\RequirePackage{fix-cm}
\documentclass[smallextended]{svjour3}       
\smartqed  
\usepackage{graphicx}
\begin{document}


\title{A Two Component Model of the Vela Pulsars with Large Fractional Moment of Inertia of the Crust}
\author{P S Negi       
}


\institute{Department of Physics, \at
              Kumaun University, Nainital\\
              \email{psneginainital63@gmail.com}           
}

\date{Received: date / Accepted: date}

\maketitle

\begin{abstract}
We construct a two-component analytic model of the Vela pulsar which can reproduce the fractional crustal moment of inertia, $I_{\rm crust}/I_{\rm total} \geq 0.074$ ( where $I_{\rm 
crust}$ represents the moment of inertia of the crust and $ I_{\rm total}$ is the total moment of inertia of the star) for the mass  range $M/M_\odot \geq 1.0 - 1.96$. The models are assumed to be self-bound at the surface  density $E_a = 2\times 10^{14}\rm g{cm}^{-3}$ (like, Brecher and Caporaso \cite{Ref1}) which yields the transition density at the core-crust boundary $E_b \geq  2.105 \times 10^{14}\rm g{cm}^{-3}$ and pressure/energy-density ratio, $P_b/E_b \geq 0.00589$.The central density, $E_0$, of the models ranges from  1.263 - 1.600 $\times 10^{15}\rm g{cm}^{-3}$. The total moment of inertia, $I_{\rm total}$, and the moment of inertia of the crust component, $ I_{\rm crust}$ lie in the range  $I_{\rm 45} = $0.076 -  2.460 and 0.0056 - 0.8622 respectively  (where $I_{45}=I/10^{45}\rm g{cm}^2$. The total radii, $a$, of the models have the  values from 9.252km - 11.578km and the crustal thickness, $a_{\rm crust}$, lies in the range 0.234km - 1.551km. The mass of the crust, $M_{\rm crust}/M_\odot$, of the models varies from 0.025 - 0.263. The pressure/energy-density ratio, $P_b/E_b$, at the core-crust boundary and other physical parameters obtained  in this study for the Vela pulsar are compared with the corresponding parameters obtained in the literature on the basis of various equations of state (EOSs). That few studies available in the literature \cite{Ref2}, \cite{Ref3} which predict the fractional crustal moment of inertia about 7\% for the Vela mass as large as 1.7$M_\odot$, the present study has been able to reproduce the minimum  fractional crustal moment of inertia about 7.4\% and larger for all the values of  the mass in the range 1.0 - 1.96$M\odot$ considered for the Vela pulsar.

\end{abstract}




\section{Introduction}
The explanation of  large glitches (the sudden spin-up rates), $\Delta \nu/\nu = 10^{-6}$, observed from the Vela pulsar on the basis of vortex unpinning model of glitch generation had imposed the constraint on the fractional crustal moment of inertia, $I_{\rm crust}/I_{\rm total} \geq 0.016$ \cite{Ref4}. However the recently discussed crustal `entrainment' effect  has modified the above constraint to $I_{\rm crust}/I_{\rm total} \geq 0.07$ \cite{Ref5}, \cite{Ref6}. By imposing the recent constraint on fractional crustal moment of inertia, various neutron star (NS) models on the basis of different EOSs have been studied under slow rotation approximation. That various studies carried out in this regard may fulfill the recent constraint on the fractional crustal moment of inertia, but yield the mass of the Vela pulsar $\leq 1 M_\odot$. Since it is generally believed that the Vela pulsar is formed in a Type II supernova (SN II) explosion, therefore its mass should be significantly larger than about 1$M_\odot$. This is what is called the `glitch crises'. Therefore, it is argued in the literature that the  crust is not enough to explain the large glitches of the Vela pulsar on the basis of vortex unpinning model so that  the core superfluid must also participate in the glitches \cite{Ref5}, \cite{Ref6},\cite{Ref7}.

 By using NS observations, Steiner et al \cite{Ref3} obtained a crustal fraction of the moment of inertia as large as 10\% for a mass $M = 1.4M_\odot$ to explain the glitches in the Vela pulsar even with a large amount of superfluid `entrainment'  \cite{Ref5}, \cite{Ref6}  which may be reduced to a maximum value of about 7\% for a mass of $1.7M_\odot$. Delsate et al \cite{Ref8} have calculated the crustal moment of inertia of glitching pulsars for different unified dense matter EOSs in order to explain the large glitches observed  in the case of the Vela pulsar. Piekarewicz et al \cite{Ref2} computed the fractional moment of inertia of neutron stars of various masses using a representative set of relativistic mean-field models. Their study suggested that the crustal moment of inertia is sensitive to the transition pressure at the crust-core interface. By maximizing the transition pressure and providing the description of nuclear observables, they have been able to obtain fractional moments of inertia as large as 7\% for neutron stars with masses below 1.6 $M_\odot$. In a recent study, Li et al\cite{Ref9} have predicted the internal structure of Vela pulsar on the basis of modern EOSs, namely  BCPM, BSk21, BSk20 and Av18* ( Li et al\cite{Ref9}; and references therein) by considering the mass range, 1.0$M_\odot$ - 2.0$M_\odot$ for the Vela pulsar. Since all four EOSs could support NS mass as large as 2$ M_\odot$, thus fulfill the constraints imposed from the two recent precisely measured  largest pulsar masses around 2$M_\odot$   \cite{Ref10}, \cite{Ref11}. Their study show that the larger value of transition pressure at core-crust boundary does not result the larger value of fractional crustal moment of inertia and they have obtained the maximum fractional crustal moment of inertia about 6.28\% for a mass around 1$M_\odot$.

Keeping in mind the various constraints  mentioned above, imposed by the observations as well as by the theory, we propose in the present study an analytic core-crust model of NS based on the study of Negi et al \cite {Ref12} and Negi\cite{Ref13}.  Together with other global properties,  
the model may reproduce the required fractional crustal moment of inertia larger than about 7\% for the realistic mass range adopted for the Vela pulsar in the literature. Furthermore, the observational constraint imposed by the  largest measured pulsar  masses around 2$M_\odot$ to date \cite{Ref10}, \cite{Ref11} may be satisfied by the models considered in the present study. The reason for considering analytic solutions for the core and crust regions follows as: The core of the model in the study of Negi et al\cite{Ref12} is described by Tolman's VII solution which gives the upper bound on energy-density for the largest measured
mass of a NS\cite{Ref14}. The crust of the model is described the fastest possible variation of density, as a result the compactness ratio ($m(r)/r$; in geometrized units) remains constant right from the core-crust boundary to the surface of the structure. For this model the speed of sound, $v_s$, remains finite and significantly less than the speed of light in vacuum, $c = 1$, at the surface, where the pressure becomes vanishingly small at the finite surface density.Therefore,  it seems physically plausible to  assume that the matter represents a self-bound state  at  the density of average nuclei at the surface ($E_a = 2\times 10^{14}\rm g{cm}^{-3}$ ) like, Brecher and Caporaso\cite{Ref1} who assumed the same criterion but for the extreme causal condition ($v_s = c = 1$) at the surface. In a landmark study, Kalogera \& Baym \cite{Ref15} used the two variants of Wiringa et al \cite{Ref16}(WFF88) EOS in the inner crust of NS models, corresponding to the core of the extreme causal (stiffest) EOS, and obtained the maximum mass of NS in the range $2.2 \leq M/M_\odot \leq 2.9$ for the permitted value of transition density at the core-inner crust boundary in the range  $E_{\rm ave} < E_b \leq 4E_{\rm n}$ (where $E_{\rm ave} =  2\times 10^{14}\rm g{cm}^{-3}$ represent the average nuclear matter density and $E_{\rm n} =  2.7\times 10^{14}\rm g{cm}^{-3}$  represents the nuclear saturation density). Very recently, the author\cite{Ref13} has shown that the use of the WFF88 EOS in the inner crust of a NS model is analogous to the use of Tolman's VI analytic solution in the envelope for the density range mentioned above.

  Because the core-crust boundary  plays a crucial and important role for obtaining a minimum value of the fractional crustal moment of NS models, the distinctive feature of the present study lies in the fact that unlike other models available in the literature in which the (number) density at the  core-envelope boundary, around 0.08 -  0.1 $/fm^3$ is chosen in somewhat  arbitrarily manner ( \cite{Ref2}, \cite{Ref3}, \cite{Ref9}), the boundary of the core-envelope models considered in Negi et al\cite{Ref12} was obtained analytically and  in an appropriate manner by matching of  all the four variables viz. pressure ($P$), energy density ($E$) and both of the metric parameters ($\nu$ and $\lambda$) at the core- envelope boundary without recourse to any computational method. The density at the core-crust boundary of the present models is constrained on the following basis: (i) The largest allowed mass for the models,  1.96$M_\odot$, which is as good as the observational value of largest pulsar masses known to-date \cite{Ref10}, \cite{Ref11} and (ii) The recent theoretical estimation of the minimum crustal moment of inertia $\geq 7\%$ predicted for the Vela pulsar \cite{Ref5}; \cite{Ref6}, rather than merely assuming it around nuclear saturation density. Sect.2 of the present study deals with the methodology and  relevant equations governing slow rotation of the analytic core-crust models (Negi et al\cite{Ref12}). Results of the calculations and an application of the models to the Vela pulsar and a comparison  regarding the total mass, total radius, central energy- density and the total moment of inertia of the Vela pulsar obtained on the basis of the present study with those of the  recent studies available in the literature (\cite{Ref2}, \cite{Ref3}, \cite{Ref9}; and references therein) is discussed in Sec 3.  Sect 4  summarizes  the main  results of the present study.

\section{Methodology}
In order to prevent the  analytic properties of the  model considered in the present work,  we neglect considering thin outer crust below the average nuclear density $E_{\rm ave} =  2\times 10^{14}\rm g{cm}^{-3}$. This  would also lead to insignificant changes in our results.
The density at the  core-crust  boundary, $E_b$, of the present models yields in the range $2\times 10^{14}\rm g{cm}^{-3} < E_b \leq  2.7\times 10^{14}\rm g{cm}^{-3}$  which  is the consequence of  the latest observational constraint (i) and the recent theoretical estimates (ii) (mentioned above) imposed on our models. This density range is comparable with the corresponding density  range obtained by Kalogera \& Baym\cite{Ref15} by using realistic EOS (Wringa et al\cite{Ref16} [WFF88])  in the inner crust and the extreme causal EOS in the core region of NS models. It is well known that the WFF88 EOS has been widely used in the literature for constructing the realistic NS models (see, e.g., Kalogera and Baym\cite{Ref15}; Crawford and Demianski\cite{Ref17}; and references therein).

We consider the metric corresponding to a static and spherically symmetric mass distribution in the following form
\begin{equation}
ds^2 = e^\nu dt^2 - e^\lambda dr^2 - r^2(d\theta^2 +{\rm sin}^2\theta d\phi^2)
\end{equation}
where $G = c = 1$ (we are using geometrized units) and $\nu$ and $\lambda$ are functions of `r' alone.
The relevant equations governing the core ($0\leq r \leq b$) and  the crust ($b\leq r \leq a$) are described respectively by Tolman's  VII and VI solutions of the Einstein's field equations for the metric (Eq.1) and are available in 
Negi et al\cite{Ref12}. However, some relations relevant to the present study are redefined as:

$u \equiv M/a$ is called the compactness parameter which is defined as the mass to size (radius) ratio of entire configuration; where
 the mass, $M = \int_{0}^{a} 4\pi E r^2dr$; and $y = r/a$ is called the radial coordinate measured in units of configuration size.

$Q(\equiv K^2/a^2)$ is defined as the compressibility factor, $K$ is a constant appearing in the Tolman's type VII solution. The matching of various parameters at the core-envelope boundary yields

$b^2/a^2Q = 5/6$, thus $(b/a)$ represents the boundary, $y_b$, of the core-envelope models and `b' represents the core radius. The analytic  relation between central to boundary density, $E_0/E_b$, and central to surface density are given by:
$(E_0/E_b) = 6$; $(E_0/E_a) = 7.2/Q$.

For slowly rotating spherical objects like the Vela pulsar (rotation velocity about 70 rad sec$^{-1}$), the macroscopic parameters of the stars are affected by first-order rotation effects. Since the Lense-Thirring frame dragging effect is a first-order effect which turns out to be about 1-2 \%  for the Vela pulsar. This effect is taken into consideration for calculating the moment of inertia of the models in the following manner [ note that the effects like deformation from spherical symmetry and mass shifts represent second order effects which are significant only for the millisecond pulsars and for the Vela pulsar the second order effects turn out to be about 10$^{-4}$ or even lower(see, e.g., Arnet and Bowers\cite{Ref18}; Crawford and Demianski\cite{Ref17}). Therefore these effects can be safely ignored in the present study which deals with the macroscopic parameters of slowly rotating configurations]:

For slowly rotating structures a perturbation solution for the metric (Eq. 1) yields (Borner\cite{Ref19}; Chandrasekhar \& Miller\cite{Ref20}; Irvine\cite{Ref21})
\begin{equation}
(d/dr)[F(d\chi/dr)] = G \chi
\end{equation}
where
\begin{equation}
F = e^{-(\nu+\lambda)/2}r^4
\end{equation}
\begin{equation}
G = 16\pi(P + E) e^{(\lambda - \nu)/2}r^4
\end{equation}
\begin{equation}
\chi = \omega - \Omega
\end{equation}
 $\omega$ being the angular velocity of structure and $\Omega$ the drag of the local inertial tetrad (Hartle and Sharp\cite{Ref22}; Hartle\cite{Ref23}); known as Lense-Thirring effect.

Substituting $F(d\chi/dr) = \phi$ in Eq. (2) we get
\begin{equation}
d\chi/dr = \phi/F
\end{equation}
and
\begin{equation}
d\phi/dr = G\chi
\end{equation}
Substituting $r = ay$, $ d r = a dy$  in Eqs. (6) and (7) we get
$
d\chi/dy = (a\phi/F) = (a\phi/a^4f)
$
and
$(d\phi/dy) = G\chi a = a^2g\chi a
$
with $f =  e^{-(\nu+\lambda)/2}y^4$ and $g = 2(8\pi Pa^2 + 8\pi Ea^2)e^\lambda f$
or
\begin{equation}
[d/dy](\phi/a^3) =  g\chi
\end{equation}
and
\begin{equation}
d\chi/dy = (\phi/a^3)/f
\end{equation}
Substituting $\phi/a^3 = \psi$ in Eqs. (8) and (9) we have
\begin{equation}
d\psi/dy = g\chi
\end{equation}
\begin{equation}
d\chi/dy = \psi/f
\end{equation}
Eqs.(10) and (11) provide a set of two first order coupled differential equations which may be solved numerically by using the standard Runge - Kutta method with boundary conditions
\begin{equation}
\chi_{y=0} = 1; (d\chi/dy)_{y=0} = 0
\end{equation}
Integrating from the  centre ($y = 0$) to the core-crust boundary ($y = b/a$) by using Tolman's  VII solution and from the  boundary ($y = b/a$) to the surface  ($y = 1$, i.e. $r = a$ and $P = 0$)  by using Tolman's type VI solution.We find that at the surface
\begin{equation}
\omega = \chi_a + (\phi_a/3a^3) = \chi_a + (\psi_a/3)
\end{equation}
Drag is given by the equation
\begin{equation}
\Omega = \omega - \chi; \rm {or} (\Omega/\omega) = 1 - (\chi/\omega)
\end{equation}
We define central drag as
\begin{equation}
 (\Omega/\omega)_0 = 1 - (1/\omega); \chi_0 = 1
\end{equation}
Thus the surface drag is given by
\begin{equation}
 (\Omega/\omega)_a = 1 - (\chi_a/\omega)
\end{equation}
The moment of Inertia, $I$, of the configuration is given by
\begin{equation}
  I = (\phi_a/6\omega) = (\psi_a a^3/6\omega)
\end{equation}

Since the present models deal with the construction of NS models on the basis of analytic solutions of Einstein's field equations,  an approximate but very precise empirical formula may be used as an alternatively to the method mentioned above  for obtaining the moment of inertia of the configurations under slow rotation. The empirical formula which is based on the numerical results obtained for the set of a large number of EOSs of dense nuclear matter yields in the following form (Bejger and Haensel\cite{Ref24})
\begin{equation}
I = (2/5)(1 + x)Ma^2
\end{equation}
where $x$ is the dimensionless compactness parameter measured in units of $M_\odot$(in\, km)/km, that is
\begin{equation}
x = (M/a)/M_\odot (\rm {in km})\rm {km}^{-1})
\end{equation}



\section{Results and application of the models to the Vela Pulsar}
Since the Vela pulsar is supposed to be the remnant of the type II supernova explosion, its mass must be higher than about 1$M_\odot$. The recent studies indicate that the mass of the Vela pulsar may be around 1.4 - 1.7$M_\odot$ \cite{Ref2},\cite{Ref3}, we have, therefore adopted the mass range, $M = 1M_\odot - 1.96M_\odot$, for the Vela pulsar. Since the models considered in the present study are applicable for a $u$ value $\leq 0.25$, it follows that for an assigned value of the surface density, $E_a = 2\times 10^{14}\rm g{cm}^{-3}$ (considered in the present study), the total mass depends only on the compactness ratio, $u$, of the whole configuration. Thus the maximum mass $M = 1.96 M_\odot$ corresponds to the maximum allowed value of $u = 0.25$. The compressibility factor $0\leq Q \leq 1.2$ defines the core-crust boundary, $y_b = (b/a)$, of the models independent of the value of $u$. It is, therefore, possible to fix a common core-crust boundary (by assigning a suitable $Q$ value) for all the model masses considered in the present study which thereafter yields a mass- independent value of fractional crustal moment of inertia. Because various studies available in the literature indicate that the majority of superfluid neutrons lie near the nuclear saturation density, $E_{\rm nm}$, corresponding to the inner crust\cite{Ref2},\cite{Ref3}, \cite{Ref9} which act as a reservoir for the moment of inertia of the inner crust. Therefore, it seems plausible to choose the transition density at the core-crust boundary in the range $E_a < E_b \leq E_ {\rm nm}$. We , therefore, choose the four values of $Q = 1.14$, 1.1, 1.0 and 0.9 from the allowed values of $Q$ in the range mentioned above which yield the transition density in the range $2.105\times 10^{14}\rm g{cm}^{-3} \leq $$ E_b$$ \leq  2.7\times 10^{14}\rm g{cm}^{-3}$. The range of transition density thus obtained is also supported by the study of Kalogera and Baym\cite{Ref15} who used the WFF88 EOS in the inner crust of their NS models below the nuclear saturation density, because of the fact that the results obtained by WFF represent the best microscopic EOS for dense matter constrained by nucleon-nucleon scattering data (\cite{Ref15}; and the references therein).

As shown in Table 1 - 4, for the transition density, $E_b$, in the range: (2.14 - 2.7)  $\times 10^{14}\rm g{cm}^{-3}$, the fractional crustal moment of inertia, $I_{\rm cr}/I_a$, yields in the range: 0.074 - 0.350, corresponding to the selected mass range: 1.0$M_\odot - 1.96M_\odot$ for the Vela pulsar. The corresponding total radius, $a$, for the said mass range lies between 9.25 km - 11.58 km and the total moment of inertia, $I_a$, ranges from 0.76 - 2.46  $\times 10^{45} {\rm g cm^2}$. For the said mass range, the pressure to energy-density ratio at the core-crust boundary, $P_b/E_b$, turns out to be in the range: 0.0058 - 0.0671, and the central energy-density, $E_0$, varies from 1.2 - 1.6$\times 10^{15}\rm g{cm}^{-3}$. The calculated values of compactness parameter, $u$, of the models yield in the range: 0.1596 - 0.24999 while the crustal mass, $M_{\rm cr}$, and the crustal radius (thickness), $a_{\rm cr}$, lie in the range: 0.02532 - 0.2627$M_\odot$ and 0.2343 - 1.5513 km respectively.

In order to compare our models with those of the models which are considered to be successful in explaining the large glitches of the Vela pulsar (i.e., the models predict large fractional crustal moment of inertia about 7\% or higher for the model mass significantly higher than about 1$M_\odot$), we consider the models presented by Piekarewicz et al\cite{Ref2} and Stainer et al\cite{Ref3}. Among the models presented by Piekarewicz et al\cite{Ref2} on the basis of  relativistic mean field theory, the models so called NL$_3{\rm max}$ and TF$_c{\rm max}$  yield the total radius and the total moment of inertia of the configurations in the range: 13.5 - 16.17 km and 1.755 - 1.866 $\times 10^{45} {\rm g cm^2}$  for the model mass of 1.4$M_\odot$.  The fractional crustal moment of inertia of the said  models turn out to be about 7.38 - 9.865\% and pressure to energy-density ratio at the core-crust boundary, $P_t/E_t$ (in their notation), lies between 0.007 - 0.009.  Stainer et al\cite{Ref3}have used the observational data set for (i) mass-radius measurements; (ii) including the moment of inertia of a 1.4$M_\odot$ NS together with (i) and (iii) the data set that includes only the moment of inertia observations excluding the data set (i) and constructed the models for Vela pulsar by employing the EOS dense matter on the basis of so called the GCR models A and GCR models C (see, e.g., Stainer et al\cite{Ref3} and references therein). For a fixed value of moment of inertia about 1.791 $\times 10^{45} {\rm g cm^2}$ for a model mass of 1.4$M_\odot$, the GCR model A of Stainer et al\cite{Ref3}yields the maximum fractional crustal moment of inertia abou 10\%. The corresponding radius of the configuration lies between 12.39 - 14.47 km and the  $P_t/E_t$ ratio of the models lie between 0.0047 - 0.0087.   The corresponding total radius and moment of inertia for a model mass of 1.4$M_\odot$, our study yields the values about 10.35 km and 1.36 $\times 10^{45} {\rm g cm^2}$ respectively, which are less than the values obtained by Piekarewicz et al\cite{Ref2} and Stainer et al\cite{Ref3}. However, our models yield the values of fractional crustal moment of inertia and the ratio of pressure to energy-density at the core-crust boundary $\geq 7.4\%$ and $\geq 0.008$ for the model mass of 1.4$M_\odot$ as shown in Table 1 - 4, which are higher than the values obtained in the study of  Piekarewicz et al\cite{Ref2} and Stainer et al\cite{Ref3}.

Apart from the models discussed by  Piekarewicz et al\cite{Ref2} and Stainer et al\cite{Ref3}, we also compare our results with those of the study of  Li et al\cite{Ref9}, who have predicted the internal structure of Vela pulsar on the basis of the  EOSs, so called the BCPM, BSk21, BSk20 and Av18* for the Vela pulsar. For this purpose, they have adopted the slow-rotation approximation, as we have made in the present paper, and solved the equations of the stellar structure for its mass, radius, and moment of inertia from the said EOSs. Four unified NS EOSs (BCPM, BSk21, BSk20and Av18*) are employed in their calculations.   For the said mass range, 1.0 - 2.0$M_\odot$, the ratio of pressure to energy-density, $P_t/E_t$ at the core-crust boundary corresponding to all four EOSs considered by Li et al\cite{Ref9} varies between 0.0035 - 0.0056 and the central energy-density varies from 0.52 - 1.88  $\times 10^{15}\rm g{cm}^{-3}$. Among their results obtained for the said  EOSs mentioned here, the results of the present study (except the fractional crustal moment of inertia which we have obtained higher than about 7.4\% for all our models) are found to be in good agreement with those of the results of Li et al\cite{Ref9} which have been obtained on the basis of BSk20 EOS for the masses higher than about 1.8$M_\odot$. By employing BSk20 EOS for the selected mass range $1.0M_\odot - 2M_\odot$ they have obtained the total radius, total moment of inertia and the fractional crustal moment of inertia in the range: 11.26 - 11.8 km, 0.894 - 2.552 $\times 10^{45} {\rm g cm^2}$  and 5.33 - 1.0\%. However for the mass range between 1.8 - 2.0$M_\odot$, the total radius and the moment of inertia for the models obtained by Li et al\cite{Ref9} yield the values in the range 11.58 - 11.26 km and 2.17 - 2.55$\times 10^{45} {\rm g cm^2}$ respectively, which are in good agreement with the results for the present study as shown in Table 1.


\section{Summary}

The modified minimum value of  fractional crustal moment  of inertia ($I_{\rm cr}/I_a$), due to  the crustal `entrainment' effect, required to explain the large Vela glitches has increased to a minimum value of about 7\%. The core-crust models obtained in this study yield this minimum required value of $I_{\rm cr}/I_a$ for the transition (energy) density, $E_b = 2.105\times 10^{14}\rm g{cm}^{-3}$, corresponding to all the model masses permissible in the range $1.0M_\odot \leq M \leq 1.96M_\odot$.  The corresponding ratio of pressure to energy--density, $P_b/E_b$, at the transition boundary is obtained in the range 0.0059 - 0.0126 for the model masses considered in this study. On the basis of modern EOSs of dense nuclear matter, it is discussed in the literature that the neutron superfluid is dominated near the nuclear saturation density. If we accept this argument as a fact and assume that the majority of neutron superfluid lies in the density range $2\times 10^{14}\rm gcm^{-3}$ - $2.7 \times 10^{14}\rm gcm^{-3}$ and choose the transition density in the range $2.105\times 10^{14}\rm gcm^{-3} < E_b \leq 2.7\times 10^{14}\rm gcm^{-3}$, the fractional crustal moment of inertia increases from the minimum value about 7.4\% to  maximum value up to 35.0\% for all the values of masses in the range $1.0M_\odot \leq M \leq 1.96M_\odot$. The  total moment of inertia of the Vela pulsar on the basis of the present study ranges from the value about $ 0.76 \times 10^{45} {\rm g cm^2}$ to  $2.46 \times 10^{45} {\rm g cm^2}$ for the assigned values of masses $M = 1.0 - 1.96M_\odot$ respectively. The total radius and the  central energy-density corresponding to the said masses are obtained as 9.25 - 11.58 km and $1.26 - 1.6 \times 10^{15}\rm gcm^{-3}$ respectively.

It is argued in the literature that the higher value of transition pressure at the core-crust boundary may lead to a higher value of crustal moment of inertia (see, e.g., \cite{Ref2}, \cite{Ref3}). However,  the argument opposite to this conclusion is also available \cite{Ref9}. But the present study clearly indicates that the higher value of the {\em ratio of pressure to energy- density} always leads to a higher value of the crustal moment of inertia. Furthermore, the present study indicates that the use of the Tolman's VII solution in the core and  the use of the Tolma's  VI solution in the crust in the two-component vortex unpinning model can provide the similar results as those obtained by Li et al\cite{Ref9} by using the BSk20 EOS together with the  higher values of fractional crustal moment of inertia required to explain the large glitches observed in the Vela pulsar,  if the Vela mass is higher than about 1.8$M_\odot$. The models based on BSk20 EOS, however,  could not provide the fractional crustal moment of inertia higher than about 1.49\% if the Vela mass is higher than1.8$M_\odot$ . The use of the  analytic solutions as an alternate to the realistic NS models lies in the fact that the various physical parameters like - pressure, energy-density, and both of the metric parameters ($\nu$ and $\lambda$)  have a direct dependence on the radial co-ordinate, $r$, which is, in general, unavailable in the NS models composed of EOSs. Thus, the internal structure of NSs can be explored in a  more simple and elegant manner on the basis of the two- component analytic models as presented in the present study.

\begin{table}
\caption{The Core-Crust Boundary, $(b/a) = 0.97468$, Total Radius, $a$(km), Core Radius, $b$(km), Pressure to energy-density ratio at the Core-Crust Boundary, $P_b/E_b$, Crust Mass  $M_{\rm cr}/M_\odot$, Crust thickness $a_{\rm cr}$(km), Fractional Crustal Moment of Inertia,  $I_{\rm cr}/I_a$ and the Total Moment of Inertia $I_{a,45} = I_a/10^{45}(\rm gcm^2)$ of the Vela pulsar for the mass range $M = 1.0M_\odot - 1.96M_\odot$. The  value of compressibility factor is obtained  as $Q = 1.14$ so that the minimum Fractional Crustal Moment of Inertia turns out to be larger than about 7.4\% for all values of masses in the range mentioned above. The value of Surface Density, $E_a$, is assumed to be the average nuclear density,  $2\times 10^{14}\rm gcm^{-3}$ (like Brecher \& Caporaso \cite {Ref1}) for all the models and the calculated value of compactness parameter $u$ corresponding to each member of the the said mass range is also shown in the Table 1. For the fixed value of  $E_a$, the energy-density at the Core-Crust boundary $E_b$ turns out to be $2.105\times 10^{14}\rm gcm^{-3}$ and the  Central  Energy-Density $E_0$  corresponds to a value $1.263\times 10^{15}\rm gcm^{-3}$.}
\label{tab:1}       
\begin{tabular}{llllllllll}
\hline\noalign{\smallskip}
$M/M_\odot$ & $a$(km) &$ u$ &$P_ b/E_b$ & $b$(km)  & $M_b/M_\odot$ & $M_{\rm cr}/M_\odot$  & $a_{\rm cr}$(km) & $I_{\rm cr}/I_a$ & $I_{a,45}$ \\
\noalign{\smallskip}\hline\noalign{\smallskip}
1.0 & 9.25221& 0.15964 & 0.0059 & 9.01794  &0.97468 & 0.02532  & 0.23427 & 0.0740& 0.7588 \\
1.2 & 9.83194& 0.18027 & 0.0071 & 9.58300  & 1.16962 & 0.03038  & 0.24895 & 0.0740 & 1.0413  \\
1.4 & 10.35035&0.19978 & 0.0084 & 10.08827 & 1.36455 & 0.03545  & 0.26208 & 0.0740& 1.3621 \\
1.6 & 10.82145 &0.21838 & 0.0098 & 10.54744  & 1.55948 & 0.04051  & 0.27400 & 0.0740& 1.7206 \\
1.8& 11.25476 & 0.23622& 0.0113 & 10.96978 & 1.75442 & 0.04558 & 0.28497& 0.0740 & 2.1157 \\
1.96 & 11.57882& 0.24999 &0.0127 & 11.28564& 1.91015& 0.04985 & 0.29318 & 0.0741 & 2.4580 \\
\noalign{\smallskip}\hline\noalign{\smallskip}
\end{tabular}
\end{table}

\begin{table}
\caption{The Core-Crust Boundary, $(b/a) = 0.95743$, Total Radius, $a$(km), Core Radius, $b$(km), Pressure to energy-density ratio at the Core-Crust Boundary, $P_b/E_b$, Crust Mass  $M_{\rm cr}/M_\odot$, Crust thickness $a_{\rm cr}$(km), Fractional Crustal Moment of Inertia,  $I_{\rm cr}/I_a$ and the Total Moment of Inertia $I_{a,45} = I_a/10^{45}(\rm gcm^2)$ of the Vela pulsar for the mass range $M = 1.0M_\odot - 1.96M_\odot$. The  value of compressibility factor is obtained  as $Q = 1.10$ so that the minimum Fractional Crustal Moment of Inertia turns out to be larger than about 12\% for all values of masses in the range mentioned above. The value of Surface Density, $E_a$, is assumed to be the average nuclear density,  $2\times 10^{14}\rm gcm^{-3}$ (like Brecher \& Caporaso \cite {Ref1}) for all the models and the calculated value of compactness parameter $u$ corresponding to each member of the the said mass range is also shown in the Table 2. For the fixed value of  $E_a$, the energy-density at the Core-Crust boundary $E_b$ turns out to be $2.180\times 10^{14}\rm gcm^{-3}$ and the  Central  Energy-Density $E_0$  corresponds to a value $1.309\times 10^{15}\rm gcm^{-3}$.}
\label{tab:1}       
\begin{tabular}{llllllllll}
\hline\noalign{\smallskip}
$M/M_\odot$ & $a$(km) &$ u$ &$P_ b/E_b$ & $b$(km)  & $M_b/M_\odot$ & $M_{\rm cr}/M_\odot$  & $a_{\rm cr}$(km) & $I_{\rm cr}/I_a$ & $I_{a,45}$ \\
\noalign{\smallskip}\hline\noalign{\smallskip}
1.0 & 9.25221& 0.15964 & 0.0098 & 8.85832  &0.95743 & 0.04257  & 0.39389 & 0.1223& 0.7588 \\
1.2 & 9.83194& 0.18027 & 0.0119 & 9.41337 & 1.14891 & 0.05109  & 0.41857 & 0.1223& 1.0413  \\
1.4 & 10.35035&0.19978 & 0.0140 & 9.90970& 1.34040& 0.05960 & 0.44064 & 0.12233& 1.3621 \\
1.6 & 10.82145 &0.21838 & 0.0164 & 10.36075  & 1.53188 & 0.06812  & 0.46070 & 0.1223 & 1.7206 \\
1.8& 11.25476 & 0.23622& 0.0190 & 10.77562 & 1.72337& 0.07663 & 0.47915& 0.1223 & 2.1157 \\
1.96 & 11.57882& 0.24999 &0.0213 & 11.0859& 1.87642& 0.08358 & 0.49295 & 0.1223 & 2.4580 \\
\noalign{\smallskip}\hline\noalign{\smallskip}
\end{tabular}
\end{table}

\begin{table}
\caption{The Core-Crust Boundary, $(b/a) = 0.91287$, Total Radius, $a$(km), Core Radius, $b$(km), Pressure to energy-density ratio at the Core-Crust Boundary, $P_b/E_b$, Crust Mass  $M_{\rm cr}/M_\odot$, Crust thickness $a_{\rm cr}$(km), Fractional Crustal Moment of Inertia,  $I_{\rm cr}/I_a$ and the Total Moment of Inertia $I_{a,45} = I_a/10^{45}(\rm gcm^2)$ of the Vela pulsar for the mass range $M = 1.0M_\odot - 1.96M_\odot$. The  value of compressibility factor is obtained  as $Q = 1.0$ so that the minimum Fractional Crustal Moment of Inertia turns out to be larger than about 23.9\% for all values of masses in the range mentioned above. The value of Surface Density, $E_a$, is assumed to be the average nuclear density,  $2\times 10^{14}\rm gcm^{-3}$ ( like Brecher \& Caporaso \cite {Ref1}) for all the models and the calculated value of compactness parameter $u$ corresponding to each member of the the said mass range is also shown in the Table 3. For the fixed value of  $E_a$, the energy-density at the Core-Crust boundary $E_b$ turns out to be $2.4\times 10^{14}\rm gcm^{-3}$ and the  Central  Energy-Density $E_0$  corresponds to a value $1.44\times 10^{15}\rm gcm^{-3}$.}
\label{tab:1}       
\begin{tabular}{llllllllll}
\hline\noalign{\smallskip}
$M/M_\odot$ & $a$(km) &$ u$ &$P_ b/E_b$ & $b$(km)  & $M_b/M_\odot$ & $M_{\rm cr}/M_\odot$  & $a_{\rm cr}$(km) & $I_{\rm cr}/I_a$ & $I_{a,45}$ \\
\noalign{\smallskip}\hline\noalign{\smallskip}
1.0 & 9.25221& 0.15964 & 0.0199 & 8.44607  &0.91287 & 0.08713  & 0.80614 & 0.2393& 0.7588 \\
1.2 & 9.83194& 0.18027 & 0.0241 & 8.97529 & 1.09544 & 0.10455  & 0.85665 & 0.2393& 1.0413  \\
1.4 & 10.35035&0.19978 & 0.0286 & 9.44853& 1.27802& 0.12198 & 0.90182 & 0.2393& 1.3621 \\
1.6 & 10.82145 &0.21838 & 0.0334 & 9.87859 & 1.46059 & 0.13941 & 0.94286 & 0.2393 & 1.7206 \\
1.8& 11.25476 & 0.23622& 0.0388 & 10.27415 & 1.64317& 0.15683 & 0.98062& 0.2393& 2.1157 \\
1.96 & 11.57882& 0.24999 &0.0436 & 10.56997& 1.78909& 0.17091 & 1.00885 & 0.2393 & 2.4580 \\
\noalign{\smallskip}\hline\noalign{\smallskip}
\end{tabular}
\end{table}

\begin{table}
\caption{The Core-Crust Boundary, $(b/a) = 0.866025$, Total Radius, $a$(km), Core Radius, $b$(km), Pressure to energy-density ratio at the Core-Crust Boundary, $P_b/E_b$, Crust Mass  $M_{\rm cr}/M_\odot$, Crust thickness $a_{\rm cr}$(km), Fractional Crustal Moment of Inertia,  $I_{\rm cr}/I_a$ and the Total Moment of Inertia $I_{a,45} = I_a/10^{45}(\rm gcm^2)$ of the Vela pulsar for the mass range $M = 1.0M_\odot - 1.96M_\odot$. The  value of compressibility factor is obtained  as $Q = 0.9$ so that the minimum Fractional Crustal Moment of Inertia turns out to be larger than about 35\% for all values of masses in the range mentioned above. The value of Surface Density, $E_a$, is assumed to be the average nuclear density,  $2\times 10^{14}\rm gcm^{-3}$ (like Brecher \& Caporaso \cite {Ref1}) for all the models and the calculated value of compactness parameter $u$ corresponding to each member of the the said mass range is also shown in the Table 4. For the fixed value of  $E_a$, the energy-density at the Core-Crust boundary $E_b$ turns out to be $2.7\times 10^{14}\rm gcm^{-3}$ and the  Central  Energy-Density $E_0$  corresponds to a value $1.6\times 10^{15}\rm gcm^{-3}$.}
\label{tab:1}       
\begin{tabular}{llllllllll}
\hline\noalign{\smallskip}
$M/M_\odot$ & $a$(km) &$ u$ &$P_ b/E_b$ & $b$(km)  & $M_b/M_\odot$ & $M_{\rm cr}/M_\odot$  & $a_{\rm cr}$(km) & $I_{\rm cr}/I_a$ & $I_{a,45}$ \\
\noalign{\smallskip}\hline\noalign{\smallskip}
1.0 & 9.25221& 0.15964 & 0.0303 & 8.01265  &0.86602 & 0.13397 & 1.23956 & 0.3505& 0.7588 \\
1.2 & 9.83194& 0.18027 & 0.0366 & 8.51471& 1.03923& 0.16077& 1.31723 & 0.3505& 1.0413  \\
1.4 & 10.35035&0.19978 & 0.0436 & 8.96366& 1.21244& 0.18756 & 1.38668 & 0.3505& 1.3621 \\
1.6 & 10.82145 &0.21838 & 0.0512 & 9.37165 & 1.38564 & 0.21436 & 1.44980 & 0.3505 & 1.7206 \\
1.8& 11.25476 & 0.23622& 0.0596 & 9.74691 & 1.55884& 0.24115& 1.50785& 0.3505& 2.1157 \\
1.96 & 11.57882& 0.24999 &0.0671 & 10.02755& 1.69728& 0.26272 & 1.55127 & 0.3505 & 2.4580 \\
\noalign{\smallskip}\hline\noalign{\smallskip}
\end{tabular}
\end{table}

\end{document}